%% file: ms.tex
\documentclass[12pt,preprint]{aastex}

\input{mycommands.tex}

\slugcomment{Accepted for publication in \apj}

\shorttitle{Resolving the Circumstellar Disk of HL Tauri}
\shortauthors{Kwon et al.}

\begin{document}

\title{Resolving the Circumstellar Disk of HL Tauri at Millimeter
Wavelengths}

\author{Woojin Kwon and Leslie W. Looney}
\affil{Department of Astronomy, University of Illinois, 1002 West Green Street, 
Urbana, IL 61801}
\email{wkwon@illinois.edu}
\and
\author{Lee G. Mundy}
\affil{Department of Astronomy, University of Maryland, College Park, MD 20742}

\begin{abstract}
We present results of high-resolution imaging toward HL Tau by the
Combined Array for Research in Millimeter-wave Astronomy (CARMA).
We have obtained \omm\ and 2.7 mm dust continua with an angular
resolution down to $0.13\arcsec$.  Through model fitting to the two
wavelength data simultaneously in Bayesian inference using a flared
viscous accretion disk model, we estimate the physical properties
of HL Tau, such as density distribution, dust opacity spectral
index, disk mass, disk size, inclination angle, position angle, and
disk thickness.  HL Tau has a circumstellar disk mass of 0.13 \msun,
a characteristic radius of 79 AU, an inclination of $40\degr$, and
a position angle of $136\degr$.  Although a thin disk model is
preferred by our two wavelength data, a thick disk model is needed
to explain the high mid- and far-infrared emission of the HL Tau
spectral energy distribution.  This could imply large dust grains
settled down on the mid plane with fine dust grains mixed with gas.
The HL Tau disk is likely gravitationally unstable and can be
fragmented between 50 and 100 AU of radius.  However, we did not
detect dust thermal continuum supporting the protoplanet candidate
claimed by a previous study using observations of the Very Large
Array at $\lambda=1.3$ cm.

\end{abstract}


\input{introObs.tex}
\input{modeling.tex}

\input{results.tex}

\input{conclude.tex}

\acknowledgments
The authors thank the CARMA staff for their dedicated work and
the anonymous referee for valuable comments.
Support for CARMA construction was derived from the states of
Illinois, Maryland, and California, the James S. McDonnell Foundation,
the Gordon and Betty Moore Foundation, the Kenneth T. and Eileen
L. Norris Foundation, the University of Chicago, the Associates of
the California Institute of Technology, and the National Science
Foundation. Ongoing CARMA development and operations are supported
by the National Science Foundation under a cooperative agreement,
and by the CARMA partner universities.
This research was supported in part by the National Science Foundation
through TeraGrid resources provided by the Purdue University
under grant number TG-AST090100. 

Facilities: \facility{CARMA}


\input{ms_kwon.bbl}
\clearpage
\input{msfigures}

\clearpage
\input{mstables}

\end{document}

%% file: mycommands.tex
\newcommand{\msun}{M$_\odot$}

\newcommand{\omm}{\mbox{$\lambda=1.3$ mm}}
\newcommand{\tmm}{\mbox{$\lambda=2.7$ mm}}

%% file: introObs.tex
\section{INTRODUCTION}
\label{sec_intro}

Young circumstellar disks are observed over a wide range of wavelengths
to study the physical conditions in the disk, 
the distribution of the gas and dust, and changes in the properties
of the disk material. The thermal emission from dust at millimeter to
sub-millimeter wavelengths is the best probe of the bulk material
distribution in the disk and the best monitor of the growth of
large grains within the disk. Due to the high angular resolutions
available with radio interferometers, continuum observations 
at these wavelengths
provide some of the best constraints on the
density distribution and the overall structure of disks during
the early era when giant planet formation is likely occurring
\citep[e.g.,][]{2010exop.book..319D}.

This paper presents sub-arcsecond imaging of the disk around
HL Tau at millimeter wavelengths with the Combined Array
for Research in Millimeter-wave Astronomy (CARMA).
HL Tau is a T Tauri star in the Taurus molecular cloud, a
nearby star forming region at a distance of 140 pc 
\citep[e.g.,][]{2004AJ....127.1029R}.  
HL Tau has been studied extensively over the
last two decades. The disk has been studied by
observations in sub/millimeter wavelength continuum using various
single dishes and interferometers: for example,
Institut de Radioastronomie Millim\'etrique (IRAM) at \omm\
\citep{beckwith1990},
Berkeley-Illinois-Maryland Association array (BIMA)
at \tmm\ \citep{1996ApJ...464L.169M,looney2000},
Very Large Array (VLA) at centimeter and millimeter wavelengths
\citep{1996ApJ...470L.117W, 2008MNRAS.391L..74G},
Owens Valley Radio Observatory (OVRO) and interferometry of
Caltech Submillimeter Observatory (CSO) and James Clerk Maxwell Telescope
(JCMT) at sub/millimeter
wavelengths \citep{lay1997}, and JCMT at submillimeter wavelengths
\citep[]{chandler2000}.  In particular, \citet{1996ApJ...464L.169M}
found physical properties such as density and temperature distribution,
disk mass, outer radius, and inclination angle, through power-law
disk model fitting to sub-arcsecond ($1\farcs32\times0\farcs48$)
angular resolution data of BIMA at \tmm, and
\citet{lay1997} studied the HL Tau disk properties using
Bayesian inference for the first time.  \citet{2002ApJ...581..357K}
also studied HL Tau by modeling of a viscous accretion disk as well
as a power-law disk.  In addition to the disk structure, HL Tau
has an optical
jet and molecular bipolar outflow \citep[e.g.,][]{1990A&A...232...37M}.
\citet{2000ApJ...540..362W} showed that HL Tau is located in a wall
of a bubble structure using $^{13}$CO observations of the BIMA
array and the NRAO 12 m telescope and  
\citet{2007ApJS..169..328R} reported that HL Tau has an envelope
component through modeling to spectral energy distribution over optical,
infrared, and submillimeter wavelengths.

\citet{2008MNRAS.391L..74G} claimed to detect a protoplanet candidate
at a projected radius of 55 AU using VLA at $\lambda=1.3$ cm 
with $0.08''$ angular resolution; the position is 
coincident with a secondary peak in previous \omm\ data 
\citep{2004IAUS..213...59W}.
\citet{2009ApJ...702L.163N} discussed the possibility of planet
formation through disk fragmentation where \citet{2008MNRAS.391L..74G}
found a compact feature in HL Tau.  Based on the Toomre Q parameter
and perturbation cooling time, they argue that the compact feature could be
a result of disk fragmentation.  In contrast,
\citet{2009ApJ...693L..86C} did not detect an emission peak
corresponding to the protoplanet candidate, at $\lambda=7$ mm with
$\sim 0.05''$ angular resolution. 
 
In this paper, we map the structure of the disk around
HL Tau and reveal its properties by visibility modeling.
We briefly discuss how the CARMA data have been taken and reduced
in Section \ref{sec_obs}, and our disk modeling is presented in Section
\ref{sec_model}.  Observational and modeling results, and the implications
for the candidate protoplanet are
described and discussed in Section \ref{sec_result_discussion},
followed by conclusion in Section \ref{sec_conclusion}.

\section{OBSERVATIONS AND DATA}
\label{sec_obs}

We carried out observations toward HL Tau in the \omm\ continuum
using the A, B, and C configurations and in the 2.7 mm continuum using
the B and C configurations of CARMA \citep{woody2004}.
The datasets were taken
between 2007 November and 2009 January using the 10.4-m and 6.1-m antennas.
To check the gain calibration,
particularly for extended-array data of A and B configurations, a
test calibrator is employed to verify successful calibration.  In
addition, we used the CARMA Paired Antenna Calibration System 
for the A configuration data \citep{PACS,2010ApJ...724..493P}; 
the 3.5-m antennas continuously observe a calibrator and their data
are used to correct short atmospheric perturbations.
The improvement
of calibration was about 10--20\% in terms of image noise levels and
the size, flux, and peak intensity of the test calibrator.  
Telescope pointing during the observations were monitored using
optical pointing and radio pointing \citep{2010SPIE.7733E.115C}.
To minimize
the bias induced by flux calibration uncertainty, a common
flux calibrator (e.g., Uranus) was used to bootstrap gain calibrator
fluxes.  In addition, different array-configuration data for HL Tau
were compared at common {\it uv} distances.  Sub/millimeter
dust emission from T Tauri disks is not variable over a period of
a few years, so amplitudes at common {\it uv} positions should be
comparable even in different configuration arrays.  
We estimate that the absolute flux calibration uncertainty 
is 10\% at \omm\ and 8\% at \tmm. 

MIRIAD \citep{sault1995} was employed to calibrate and map data.
In addition to normal calibration, seeing has been corrected
using the UVCAL task for the B configuration data at $\lambda=1.3$ mm, 
which were taken under less favorable weather conditions.
Individual configuration data were calibrated separately and combined 
to make maps.  
The proper motion of HL Tau
\citep[v$_{RA}=8.0 \pm 6.0$ mas year$^{-1}$ and v$_{Dec}=-21.8 \pm 5.8$ mas year$^{-1}$; ][]{2003yCat.1289....0Z}
was compensated for the data to set the epoch positions
to 2009 January.
The sensitivity
and emphasized size scales in the maps depend on weighting schemes of
visibility data.  In order to emphasize the small structures, 
 we used Briggs robust weighting of 0 \citep{briggs1995}.

Figure \ref{fig_hltau} shows the A, B, and C configuration-combined
image at \omm\ ($\nu = 229$ GHz) with $0\farcs17 \times 0\farcs13$ 
(PA $=85.0\degr$) resolution, 24 AU by
18 AU in linear size at the distance of HL Tau. The RMS noise level in the
\omm\ map is 0.8 mJy beam$^{-1}$.
The HL Tau disk is nicely resolved;
the apparent disk size is $\sim 1\farcs5 \times 1\farcs1$ 
at the 4$\sigma$ contour
level, corresponding to 210 AU major axis.  The peak intensity at \omm\ is
33 mJy beam$^{-1}$ and the total flux in a $2 \arcsec$ box centered on
the source is $700\pm10.3$ mJy. 
A Gaussian fit to the emission yields an emission
centroid position of RA(J2000) $= 04^h~31^m~38\fs418$, 
Dec(J2000) $= +18\degr~13\arcmin~57\farcs37$.
The position angle of the major axis is $135 \degr$ East-of-North and the
inclination is $43 \degr$, based on the ratio of the major and
minor axes ($0\degr$ corresponds to a face-on disk).  
The \tmm\ ($\nu = 112$ GHz) data yield a continuum map with a 
resolution of $1\farcs01 \times 0\farcs67$ (PA $= 72.8\degr$) and an
RMS noise level of 1.1 mJy beam$^{-1}$; the
peak intensity is 64 mJy beam$^{-1}$ and the integrated flux is 
$120\pm3.8$ mJy. 
The given errors in both fluxes are statistical uncertainties;
systematic calibration uncertainties are estimated to be about 10\%.

%% file: modeling.tex
\section{DISK MODELING}
\label{sec_model}

The disk around HL Tau shows a high degree of axial symmetry and
a significant extent relative to the beam size at \omm\ wavelength.
In order to derive physical parameters for the disk,
we fit the data with a standard viscous accretion disk model
\citep{1981ARA&A..19..137P}.
This disk model has a power-law radial density distribution tapered
by an exponential function: in the case of a thin disk,
$\Sigma(R) \propto (R/R_c)^{-\gamma} \textrm{exp}[-(R/R_c)^{2-\gamma}]$,
where $R_c$ is a characteristic radius \citep[e.g.,][]{2009ApJ...700.1502A}.
For our fits, we assume a disk thickness determined by vertical hydrostatic equilibrium;
the density distribution in cylindrical coordinates is,
\begin{equation}
\rho(R,z) = \rho_0 \Big(\frac{R}{R_c}\Big)^{-p}
\textrm{exp}\Big[-\Big(\frac{R}{R_c}\Big)^{7/2-p-q/2}\Big]
\textrm{exp}\Big[-\Big(\frac{z}{H(R)}\Big)^2\Big].
\label{eq_rho}
\end{equation}
Here $H(R)$ is the scale height at a radius $R$, which is
the sound speed divided by the Keplerian angular velocity:  
$H(R)\equiv \sqrt{2}c_s/\Omega = \sqrt{2kT(R,0)R^3/GM\bar{m}}$.
The surface density power-law index can be expressed as $\gamma=-3/2+p+q/2$,
as $\Sigma(R) = \rho(R,0)H(R)/\sqrt{\pi}$.

The $q$ is the temperature power-law index of dust grains,
$T(R,0)=T_0(R_0/R)^q$.
When assuming a power-law opacity ($\kappa_\nu = \kappa_0 (\nu/\nu_0)^\beta$)
and radiative equilibrium with a central protostar,
it is expressed by $\beta$: q = 2/(4 + $\beta$) in the low optical depth
limit \citep{spitzer1978ppim}.
For calculating scale heights, we adopted the mid-plane temperature
$T_m(R,0)=T_0(R_0/R)^q$ and $q=0.43$ (corresponding
to $\beta=0.7$). 
Note that the assumed power-law index is consistent with self-consistent
temperature distributions \citep{dullemondcode}.  In contrast to the
scale height calculation, 
we utilized temperature distributions as a function of z as well
as R, which simulates self-consistent temperatures:
\begin{equation}
T(R,z) = W T_m(R,0) + (1-W) T_s(r),
\end{equation}
where $T_m(R,0)$ and $T_s(r)$ indicate mid-plane and surface temperature
distributions, respectively.  The two temperature distributions
have functions of the same power-law index of $q=0.43$: 
$T_m(R,0)=T_0(R_0/R)^q$ and $T_s(r)=T_{s0}(r_{s0}/r)^q$.  
However, note that the surface temperature
depends on $r=\sqrt{R^2+z^2}$, while the mid-plane temperature is
a function of R.  We adopt $T_{s0}=400$ K at $r_{s0}=3$ AU corresponding
to $L = 8.3~\textrm{L}_\sun$ and $W = \textrm{exp}[-z^2/2(3H(R))^2]$ resulting
in temperature distributions close to the self-consistent ones empirically,
except the very inner region showing a steep temperature gradient.
The mid-plane temperature distributions (the scale factor $T_0(R_0)$)
are most sensitive to the disk thickness so they are scaled after
checking self-consistent temperature distributions by a test run
of a specified disk thickness using the \citet{dullemondcode} code.  
The employed values are mentioned in Section \ref{sec_grainsettlement}.
This estimation of the temperature distribution may not be the best
approach but it provides a fast means to achieve modeling results
without a significant loss of accuracy.

The $\rho_0$ of the density expression in Equation (\ref{eq_rho}) can
be expressed in terms of $M_{disk}$
(total disk mass), $p$, $q$, $R_{in}$ (disk inner radius), and $R_c$
by integrating the density distribution.  We assume $\kappa_0=0.01$
cm$^2$g$^{-1}$ at $\nu_0 = 230$ GHz given by ice mantle grains
\citep{ossenkopf1994} in a size distribution with a power index of
3.5 \citep{1977ApJ...217..425M} and a gas-to-dust ratio of 100.
The inclination and position angle are also free parameters in our
modeling.

Modeling to constrain these free parameters is carried out
with various disk thickness.
For disk thickness we employ a scale height factor, $b_{height}$,
which is defined as a disk scale height with respect to 
the hydrostatic equilibrium scale height.
Therefore, $b_{height}$ smaller than 1 means that the disk appears 
thinner than the thickness of hydrostatic equilibrium in the dust 
continuum,  which implies dust grain settlement.
We adopt the central protostellar mass 
$M_{dyn}=0.55$ \msun\ estimated by protostellar evolution
tracks on the luminosity-temperature plot \citep{beckwith1990} and
by the Keplerian rotation of the disk gas \citep{1991ApJ...382L..31S}.

In summary, the parameters to be constrained for the viscous
accretion disk model are seven in total:
$p$ for a density distribution, $\beta$ for a dust opacity spectral
index, $M_{disk}$ for a disk total mass, $R_{in}$ for an inner radius,
$R_{c}$ for a characteristic radius, $\theta_i$ for an inclination, and
$PA$ for a position angle.
In addition, we investigate disk thickness by constraining the seven
parameters in various $b_{height}$ cases:
$b_{height} = 0.05$, 0.1, 0.2, 0.5, 1.0, 1.5, and 2.0.

Disk images are first calculated by numerically
solving the radiative transfer equation along the line of sight
without optically thin and Rayleigh-Jeans approximations.
The resulting model images were 
``observed'' by multiplying by three types of CARMA primary beams based on
baseline antennas (baselines of 10.1-m and 10.1-m antennas,
10.1-m and 6.4-m antennas, and 6.4-m and 6.4-m antennas), 
Fourier-transforming the image into {\it uv} visibility space, and sampling 
with the observational {\it uv} coverage.  Finally, the sampled visibility
data are compared with the observational data in Bayesian
inference \citep{gilks1996,mackay2003}.

Bayesian inference allows us to obtain the probability distribution
of disk properties--$P(m\mid D, H)$, which is a probability distribution
of a disk property parameter $m$ with given data $D$ and a given
disk model $H$:
\begin{equation}
\label{eq_bayes}
P(m \mid D, H) = \frac{P(D \mid m, H) P(m \mid H)}{P(D \mid H)}.
\end{equation}
The $P(D \mid H)$ called evidence is just a normalization factor
of the posterior $P(m \mid D, H)$ here in our modeling, which
employs only a disk model, and
we assume uniform priors $P(m \mid H)$ over parameter search ranges.
For the likelihood $P(D \mid m, H)$, we utilize Gaussian function,
as the noise of interferometric data shows a normal distribution:
$P(D \mid m, H)=\prod_i \textrm{exp}[-(D_i-M_i)^2/2\sigma_i^2]$, 
where $D_i$ is an
observational visibility, $M_i$ is a model visibility, and
$\sigma_i$ is the uncertainty of the observational visibility.
In principle, the uncertainty of interferometric visibility data is estimated
by system temperatures, bandwidths, integration time, and efficiencies of
antenna surface and correlator quantization \citep{ISRA2001}.
However, these uncertainties do not take into account
errors caused by atmospheric turbulence, pointing errors, and
overall systematic calibration uncertainties.
We found that the standard deviation of the imaginary components of 
self-calibrated gain calibrator data presents 
{\it uv} distance-dependent uncertainty for the atmospheric turbulence.
Therefore, we utilize this more-realistic uncertainty, which
is typically a few times larger than the theoretical estimate.
We point out that it is still lower limit of uncertainty, 
as we do not fully capture atmospheric variation 
for the errors caused in gain solution transfer from a calibrator to
a target and any other possible error sources, for example, 
non-identical beam patterns of even the same size antennas.

The parameter search is carried out by a Markov Chain Monte Carlo
method (Metropolis-Hastings algorithm), and in order
to deal with the multi-dimensional parameter space,
Gibbs sampling is utilized; individual parameters
are drawn based on the Metropolis-Hastings algorithm one by one,
with the other parameters fixed \citep{mackay2003}.
Our modeling has been tested with artificial disk data 
\citep{2009PhDT........18K}.

In addition to our two wavelength images\footnote{Our original data 
are visibilities
of interferometry and we carried out model fitting in the visibility
space.  However, the word ``image'' is used to indicate our data
in contrast to the spectral energy distribution data.}, 
we take into account the
spectral energy distribution (SED) of HL Tau over near-infrared to
millimeter wavelengths as well, which comes from the
literature \citep[e.g.,][]{1999ApJ...519..257M}.  
While geometrical and physical properties of a disk 
are best constrained by high angular resolution images,
a best model should also explain the SED.
The SED is affected by frequency-dependent properties
(e.g., dust opacity) as well as disk geometries.

%% file: results.tex
\section{RESULTS AND DISCUSSION}
\label{sec_result_discussion}

\subsection{Disk Thickness : Large Grain Settlement?}
\label{sec_grainsettlement}
We have obtained marginal probability distributions of the free parameters
fitting the two images in seven cases with $b_{height}$ fixed at:
$b_{height}=0.05$, 0.1, 0.2, 0.5, 1.0, 1.5, and 2.0.  Based on the
typical test runs of the seven parameters for each $b_{height}$,
we adopt $T_0=30$, 35, 41, 55, 70, 82, and 93 K at $R_0= 10$ AU for
the mid-plane temperature distributions.  In the achieved probability
distributions of parameters shown in Figure \ref{fig_diff_b}, the
case of $b_{height}=0.05$ is excluded, since the fits always hit
our disk mass limit of 0.5 \msun.  
As the bottom-right panel of Figure \ref{fig_diff_b}
presents log(posterior), the thinner disk models are more preferred
by the images.  The thinner disks have a colder mid-plane temperature
so they need more disk mass for the observed flux.  Note that, however,
the disk mass is likely overestimated in thin disks, because
the gas-to-dust ratio significantly decreases from 100 as dust grains settle
on the mid plane.  As the disk
mass goes up, optical depth increases particularly in the inner
disk region.  Therefore, the density gradient gets steeper and the
opacity spectral index $\beta$ becomes larger.  The other parameters
also change along the disk thickness ($b_{height}$) but not
significantly.

For the best fit models of each $b_{height}$ case, we calculated model
SEDs, using a Monte-Carlo radiative transfer code
\citep{dullemondcode}.  We assume a power-law opacity of the
constrained index $\beta$ for dust properties.  Figure \ref{fig_SED}
shows the cases of $b_{height}= 0.1$, 0.2, 0.5, 1.0, 1.5, and 2.0
from the bottom, distinguished particularly 
in the mid-infrared regime.  The $b_{height}=1.5$
case, which is closest to and does not overestimate the data points,
is presented by a solid line.  The open rectangular points are data
from \citet{1999ApJ...519..257M} and the solid stars mark our values
at \omm\ and \tmm.  Although the thinner disk models are preferred
by our two wavelength images of CARMA, they cannot explain 
the mid-infrared fluxes.
In contrast, the $b_{height}=1.5$ case reasonably well recovers the
SED over mid-infrared to millimeter wavelengths and fits the two
CARMA images.  The bump of the model
SEDs in near-infrared regime is the central protostar.  Since HL Tau
is located at a boundary of a wall structure \citep{2000ApJ...540..362W},
one would expect large-scale extinction of those wavelengths.

The $b_{height}$ greater than unity can be interpreted as existence
of residual envelope components.  Indeed, some previous studies
have reported envelope components of HL Tau based on SED fitting
results \citep[e.g.,][]{2007ApJS..169..328R}.  
However, note that the disk scale height
larger than the hydrostatic equilibrium value could be an
overestimate.  Since SED data have been taken with poor angular
resolution and HL Tau is located on a wall structure,
it may be likely that the wall structure also contributes to the
mid-infrared fluxes.
On the other hand, it might also be caused by
an overestimate of the central protostellar mass of HL Tau
or an underestimate of the disk mid-plane temperature.

It is noteworthy that thin disk models are preferred by the image
fitting, while thick disk models better match SED.
We do not attempt to quantitatively fit the two images and the
SED simultaneously in Bayesian inference, as it is not clear how
the significance of data should be distributed between the images
and the SED, which consist of millions of visibility data points
and tens of flux densities, respectively.  More
importantly, protoplanetary disk emission is unlikely to be
explained by a simple model with a scale height. 
In other words, disk thickness ($b_{height}$
in our modeling) can depend on dust grain sizes, as larger grains likely
settle to the mid plane faster and more compactly.  This results in
smaller scale heights for longer wavelength observations;
short wavelength observations do not probe the mid plane
due to relatively large optical depth, so a thick disk would be 
preferred.  In fact,
numerical simulations show that large grains rapidly settle into
the disk mid plane, resulting in significant differences of scale
heights along grain sizes
\citep{2009MNRAS.397...24B,2010MNRAS.403..211T}. 
This is not surprising since for a long time
dust settlement has been considered as the initial stage to form
planetesimals \citep[e.g.,][]{1973ApJ...183.1051G}.
We also found the best fitting model of only our millimeter images
with setting the $b_{height}$ as a free parameter and with a broader
disk mass range.  When the mid-plane temperature is assumed as
$T_0=30$ K at 10 AU, which is the same to the $b_{height}=0.05$
case, the marginal probability of $b_{height}$ is at maximum around
0.02 and the other parameters follow the trend shown in Figure
\ref{fig_diff_b}.  The results imply that the dust grains most
sensitive to millimeter wavelength observations have settled in a
scale height deceased by a factor of 50.

It has been noted that dust settlement decreases mid- and far-infrared 
emission \citep{2004A&A...421.1075D}, as shown in the thin disk cases
of Figure \ref{fig_SED}.  
However, HL Tau has bright mid-infrared emission; 
a disk that still has much gas and ``strong'' accretion like 
HL Tau, which has
relatively strong turbulence, can arguably keep significant
mid-infrared emission by not-settling fine dust grains mixed with gas
\citep[e.g.,][]{2009MNRAS.397...24B}.
On the other hand, we are aware that the disk thickness can also depend on
disk models (e.g., accretion disk and power-law disk), 
so it may be worthy of testing different disk models further
(Kwon et al. in preparation).

\subsection{Disk Properties Constrained by Modeling}
\label{sec_properties}

Considering the fitting results of the two wavelength images and
SED, the possibility of large grain settlement, and the bias of the
extended wall structure, it is not trivial to select the best fitting
model.  In addition, if the disk is stratified as discussed, it
might not be the best approach either.  If we simply took account
of the fitting results of the two images and SED, however, the
$b_{height}=1.5$ would be a representative of the best fitting
models, with a caveat that the disk could be stratified with large
grains settled down on the mid plane.  In addition, as dust is more
or less 1\% of the total disk mass, the $b_{height}=1.5$ case may
better present the overall disk structure of gas and dust.  We
discuss each of the model parameters focusing on the $b_{height}=1.5$
case in the following paragraphs.
On the other hand, the stratified structure may
be investigated using submillimeter observations with
high angular resolution and image fidelity, e.g., using the Atacama
Large Millimeter Array (ALMA).  Our \tmm\ image does not have a high angular
resolution comparable to \omm.  If HL Tau is apparently stratified,
a thicker disk than the one constrained by our millimeter data would
be preferred by submillimeter data.  

Figure \ref{fig_uvamp} presents {\it uv} visibilities in annulus
average with the dust opacity spectral index $\beta$ simply calculated
assuming optically thin and Rayleigh-Jeans approximation:
$\beta=\alpha-2$, where $\alpha$ is a flux density spectral index.
The overlaid solid lines are of the best fit model with $b_{height}=1.5$.
Figure \ref{fig_OMR} shows the image, model, and the residuals.
The residuals to the model at \omm\ are a mixture of
plus-minus 3$\sigma$ peaks; the residuals to the \tmm\ fit show a
4$\sigma$ peak to the northeast and a 7$\sigma$ peak to the southwest.
The \tmm\ residuals, which are formally significant, might indicate
dust opacity variation in the disk, but it is not clear due to our
poor angular resolution in the \tmm\ map.  On the other hand, the
residuals could be free-free emission from the ionized jet oriented
at this position angle and seen better at longer wavelengths 
\citep{2006A&A...446..211R}.

The probability distributions for the free parameters constrained
by our modeling with $b_{height}=1.5$ are presented 
in Figure \ref{fig_posthl} and in
Table \ref{tab_parampost}.  The parameter ranges for good fits in
the figure and the uncertainties in the table are for
our specific parameterized accretion disk models; other families
of models may also fit the data.  
In addition, the uncertainties of the figure do not
include systematic uncertainties such as absolute flux calibration
uncertainty.  The table has a column and notes for possible 
systematic uncertainties.

The disk central density power-law index p is well constrained in
the model at $p=1.06$, under the temperature
power-law index, $q=0.43$. 
Based on the fitted density power-law index, a surface density
distribution power-law index $\gamma$ of -0.22 is derived.  This
negative value does not imply a continuous increase of surface
density with radius, due to the tapering of the exponential function.
For the best fit model in Table \ref{tab_parampost}, the surface
density increases by 50\% from $R_{in}$ to 28 AU, then decreases beyond
28 AU.  While the density distribution is not straightforward to be compared
with previous studies using a power-law disk model due to
the different functional forms, a study using the accretion disk model
toward HL Tau \citep{2002ApJ...581..357K} reported a steeper surface density
gradient ($\gamma = 0.78$).  This discrepancy may be caused by the
poor angular resolution of their observations, $1\farcs2 \times 1\farcs1$.

The dust opacity spectral index $\beta$ is best fitted by a value
of 0.73.  This value is primarily determined by the ratio of the
flux measure at the two wavelengths since our model fits do not
have significant optical depths over most of the disk. As shown in
Figure \ref{fig_diff_b}, however, a thinner disk has significant
effects of optical depths resulting in a larger $\beta$.  The total
uncertainty in $\beta$ includes the statistical uncertainty, which
is captured in the figure, and the relative uncertainties in the
flux calibration between the two wavelengths. For the estimates
flux uncertainties of 8--10\% for the two wavelengths, the total
uncertainty in $\beta$ is $\pm$0.25.  
The small $\beta$ implies that dust grains have grown further
comparing protostellar systems at earlier stages, which have
$\beta\approx1$ \citep[e.g.,][]{kwon2009}.

On the other hand, \citet{2006A&A...446..211R} derived
$\beta$=1.3 in the wavelength range of 1 to 7 mm. Extension of our
$\beta$=0.73 law to 7~mm clearly overestimates the measured flux:
9.4 mJy from our model versus 3.9 mJy of disk emission estimated
by \citet{2006A&A...446..211R}. To produce the measured
7~mm flux our model needs $\beta = 1.28$ between \omm\ and 7~mm.
Note that the model with this $\beta$ produces the same \omm\ image
but it does not fit our \tmm\ flux and image.
Figure \ref{fig_kappa} shows the ratio of the mass absorption coefficient
($\kappa_\nu$) at a wavelength divided by the mass absorption
coefficient at \omm. 
A straight line in this log-log plot represents a single power
law index over the full wavelength range.  The turnover in the ratio
value could indicate a largest grain size of several millimeter in the disk
\citep{draine2006}.

The disk mass is very well determined within the context of the model:
0.135 \msun. However, the full uncertainty must include uncertainty
in the absolute flux calibration, the mass absorption coefficient, the
gas-to-dust ratio, and the
dust temperature. We have estimated the flux calibration uncertainty 
at 10\% which translates directly to a mass uncertainty. The mass
opacity coefficient could be uncertain at the factor of two level.
The total luminosity of the HL Tau system is also uncertain: estimates
range from 1 to 11 L$_\odot$. The resulting uncertainty in the dust
temperature at a fiducial radius is about 67\%.
Note that the massive disks constrained in thinner disk models
(Figure \ref{fig_diff_b}) would have larger uncertainty in
the gas-to-dust ratio, as the models imply grain settlement.
Although the mass is uncertain, we argue that this particular model 
of $b_{height}=1.5$ would have the least overestimated disk mass and 
the least uncertainty in our modeling at about a factor of two.

The inner radius is constrained by our data to be significantly
smaller than our \omm\ linear resolution of about 20 AU. The formal
model fit gives an inner radius of 2.4 AU.
Fits to the mid-infrared emission typically require disk material to
an inner radius of several stellar radii to 0.2 AU 
\citep{1999ApJ...519..257M,2008ApJS..176..184F}.
It is possible that an inner radius in
millimeter emission is tracing a surface density drop, but the material
remains optically thick at mid-infrared wavelengths throughout the
inner disk.  

The characteristic radius, which divides the regions dominant by
the power-law and the exponential density distributions, is 79 AU. It
is this exponential cutoff in Equation (\ref{eq_rho}) that sets
the outer scale for the disk in the presence of the shallow power-law
index.  The surface density falls by a factor of ten from its peak
at a distance of 125 AU, and a factor of one-hundred at 165 AU. In
the context of the accretion disk model, the transitional radius
$R_t$, the radius where the bulk flow direction changes from inward
to outward for angular momentum conservation, is then 40.3 AU:
$R_t=R_c(\delta/2)^\delta, \textrm{ where } \delta=1/(2-\gamma)$
\citep{2009ApJ...700.1502A}.  A previous study reported that $R_t$
gets larger with age \citep{2009ApJ...701..260I} but another
circumstellar disk survey did not find such a trend
\citep{2009ApJ...700.1502A}.

The disk inclination and position angle are very well determined
by the disk model: inclination of about $40\degr$ (face-on disk
$\theta_i = 0\degr$) and position angle of about $136\degr$.  In
fact, these are not varying much even over the wide range of $b_{height}$
(Figure \ref{fig_diff_b}).  The inclination is particularly consistent
with previous studies using submillimeter and millimeter wavelength
observations of a high angular resolution.  \citet{1996ApJ...464L.169M}
suggested a relatively wide range of $20\degr$--$55\degr$ and
\citet{lay1997} constrained as $42\degr\pm5\degr$.  \citet{1999ApJ...519..257M}
also reported an inclination of $47\degr$ using a detailed SED
modeling.  However, fits at infrared wavelengths have suggested
inclinations from i $=15\degr$ from mid-infrared spectral fits
\citep{2008ApJS..176..184F}, to 66--71$\degr$ from fitting to 
near-infrared images and polarization \citep{2004MNRAS.352.1347L}.
It is possible that the near- and mid-infrared determinations
of inclination are affected by large scale inhomogeneities such as 
the wall structure.

\subsection{Constraints on the Presence of the Proposed Protoplanet
and Disk Stability}

Figure \ref{fig_hltau_wres} presents the zoomed-in residual map 
overlaid on top of the image at \omm. 
Particularly the top-left panel displays a blow-up of
the \omm\ image of the disk in the region of the protoplanet candidate
\citep{2008MNRAS.391L..74G}.  The right panel shows the residuals
for the same region when the disk model is removed. The cross-hashed
circles of both panels mark the position where \citet{2008MNRAS.391L..74G}
claimed a protoplanet candidate.  \citet{2008MNRAS.391L..74G}
detected a clump of $78\pm17~\mu$Jy at $\lambda=1.3$ cm.  The clump
would be $25\pm5$ mJy at $\lambda=1.3$ mm, assuming a $\nu^{2.5}$
emission spectrum.  If the dust grains responsible for the emission were
very large so that the emission spectrum rather followed  $\nu^{2.0}$,
i.e., the hardest to detect for our observations, then
the flux at \omm\ would be $7.8\pm1.7$ mJy.  Since our observational
sensitivity at \omm\ is 0.8 mJy beam$^{-1}$, the candidate could be
detected in our observations in spite of our relatively worse
angular resolution: $0.08 \arcsec$ versus $0.13 \arcsec$.

As shown in the panels, the candidate position is within the dust thermal
emission region of the disk at \omm; however, subtraction of the
axi-symmetric disk model removes essentially all flux.  Before
model subtraction the flux at the position is $\sim$16
mJy beam$^{-1}$; after subtraction the position has a negative flux
at a level of 2$\sigma$.  Therefore, our \omm\ image provides no
support for the candidate protoplanetary condensation.
The flux level before subtraction is roughly consistent with
the value bootstrapped from the $\lambda=1.3$ cm flux, but it is part
of the large disk structure, which was not detected by their observations,
rather than a compact signal.

Although our observations do not have the protoplanet candidate signal,
the HL Tau disk could be gravitationally unstable.
Our disk model provides a surface density profile that can be used
to calculate local gravitational stability in the disk. The Toomre
Q parameter \citep{1964ApJ...139.1217T} with values smaller than 
1.5 is a good indicator for gravitationally unstable
regions.  The Q parameter versus radius, Figure \ref{fig_qplot},
was calculated with the disk temperature and density distributions
of our modeling. It shows a minimum less than or close to 1.5 in the range of
50 to 100 AU, which means that substructures can develop at these
radii by gravitational instability.  
This result is in agreement with the work by
\citet{2009ApJ...702L.163N}, which
studied the possibility of fragmentation in the HL Tau disk
based on perturbation cooling time and gravitational instability.
Considering the surface density, protostellar mass, and
the radius of our modeling in the \citet{2009ApJ...702L.163N} results, 
roughly 2 $M_{Jupiter}$ fragments are favored between 50 and 100 AU.
As discussed in Section \ref{sec_properties}, the disk mass that is
the most important property for the Q parameter calculation has a large
uncertainty.  However, since the model used here is not even the one with a
small $b_{height}$ favoring a massive disk,
the HL Tau disk is likely unstable in the outer region.

Although we have not detected a clear substructure,
the HL Tau disk is likely unstable and can be fragmented.
The positive residual at $\sim$100 AU in Figure \ref{fig_hltau_wres}
might be a hint of a dust lane.
Indeed, extrasolar planetary systems with planets at outer disk regions
have been found \citep[e.g.,][]{2008Sci...322.1345K,2008Sci...322.1348M}.
These planets of outer disk regions are thought to have formed
by gravitational instability (so-called the disk instability scenario), 
although it is little known whether and how the
protoplanets formed by gravitational instability
at the early phase of protoplanetary disks still having gas
can survive the following active migration stage
\citep[e.g.,][]{2010exop.book..319D}.
In contrast, the other major model of giant planet formation,
so-called the core accretion scenario, cannot currently explain the
giant planets at the outer disk regions due to too-large time scales
of gas accretion \citep[e.g.,][]{2010exop.book..319D}.
Regardless of giant planets being actually formed by 
disk instability or core accretion,
they should be formed in protoplanetary disks that still have
gas.  In the near future, ALMA will provide the unprecedented image
to reveal the more detailed disk structure of HL Tau.
In addition, we would be able to
distinguish the two scenarios of giant planet formation by
gas kinematics and substructure and scale height variation in 
multiple wavelength observations.

%% file: conclude.tex
\section{CONCLUSION}
\label{sec_conclusion}

We have obtained dust continuum data
of HL Tau at \omm\ and 2.7 mm with the exceptionally high
image fidelity and angular resolution up to $0.13 \arcsec$ using CARMA.
This resolution reveals 18 AU scales at the distance of HL Tau, 140 pc.
Through visibility modeling in Bayesian inference adopting the viscous
accretion disk model, we constrained the physical properties of HL Tau.
As we fit our two wavelength data
simultaneously, we also constrained the dust opacity spectral index.
In addition to the two wavelength images, we also qualitatively fit
the SED.

Modeling in Bayesian inference toward the two wavelength images
prefers a thin disk (Figure \ref{fig_diff_b}).  However, thin disk
models can not explain mid-infrared emission of SED (Figure
\ref{fig_SED}).  A thick disk is needed to produce the mid-infrared 
emission.  The discrepancy
between the fitting results of millimeter images and the SED may
indicate the settlement of particularly large grains into the mid plane
(a stratified settlement of dust grains). 
However, based on
the reasonable fitting to both the millimeter images and SED
and considering that dust is not the majority of disk mass,
we selected and further discussed the case of $b_{height}=1.5$
(Figure \ref{fig_posthl} and Table \ref{tab_parampost}), 
which possibly supports envelope 
residuals beyond the hydrostatic equilibrium thickness. 
The scale height ratio larger than unity might also be caused by
an overestimate of the HL Tau protostellar mass or an underestimate
of the disk mid-plane temperature. 
The $b_{height}=1.5$ model reveals that the disk mass is about
0.135 $M_\sun$ and the dust opacity spectral index presents that
grains have grown further, compared with objects at earlier
evolutionary stages.  The disk inclination and position angle are
well determined as $40\degr$ (face-on disk of $\theta_i = 0\degr$)
and $136\degr$, respectively.

Although a protoplanet candidate has been claimed in HL Tau,
our observations, which are more sensitive to dust thermal emission,
do not detect emission structure supporting the candidate.
However, HL Tau appears to be gravitationally unstable and
could favor fragmentation on the Jupiter mass scale between 
about 50 and 100 AU.
Further observations with a higher angular resolution and
a higher sensitivity using ALMA will be able to show 
substructures possibly formed by gravitational instability
as well as to verify the stratified settlement of dust grains
in HL Tau.

%% file: msfigures.tex
\begin{figure}
\includegraphics[angle=270,scale=0.8]{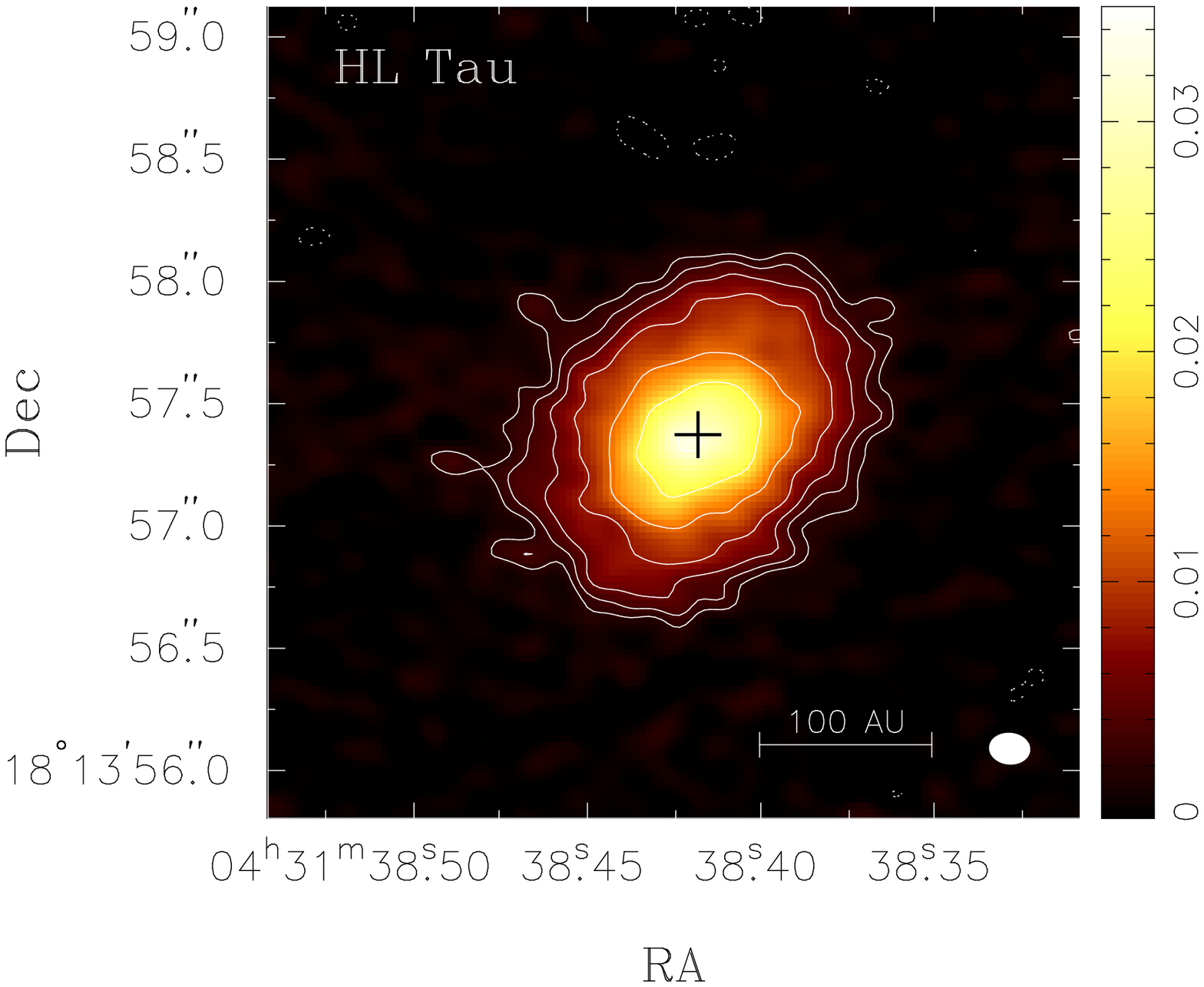}
\caption{
HL Tau in the \omm\ continuum.
The image is the combined of CARMA A, B, and C configurations, and
the synthesized beam is $0\farcs 17 \times 0\farcs 13$
($PA = 86 \degr$) corresponding to 18 AU.
The contour levels are 2.5, 4.0, 6.3, 10, 16, 25, and 40 times 
$\sigma=\pm0.8$ mJy beam$^{-1}$.
\label{fig_hltau}}
\end{figure}

\begin{figure}
\includegraphics[angle=270,scale=0.6]{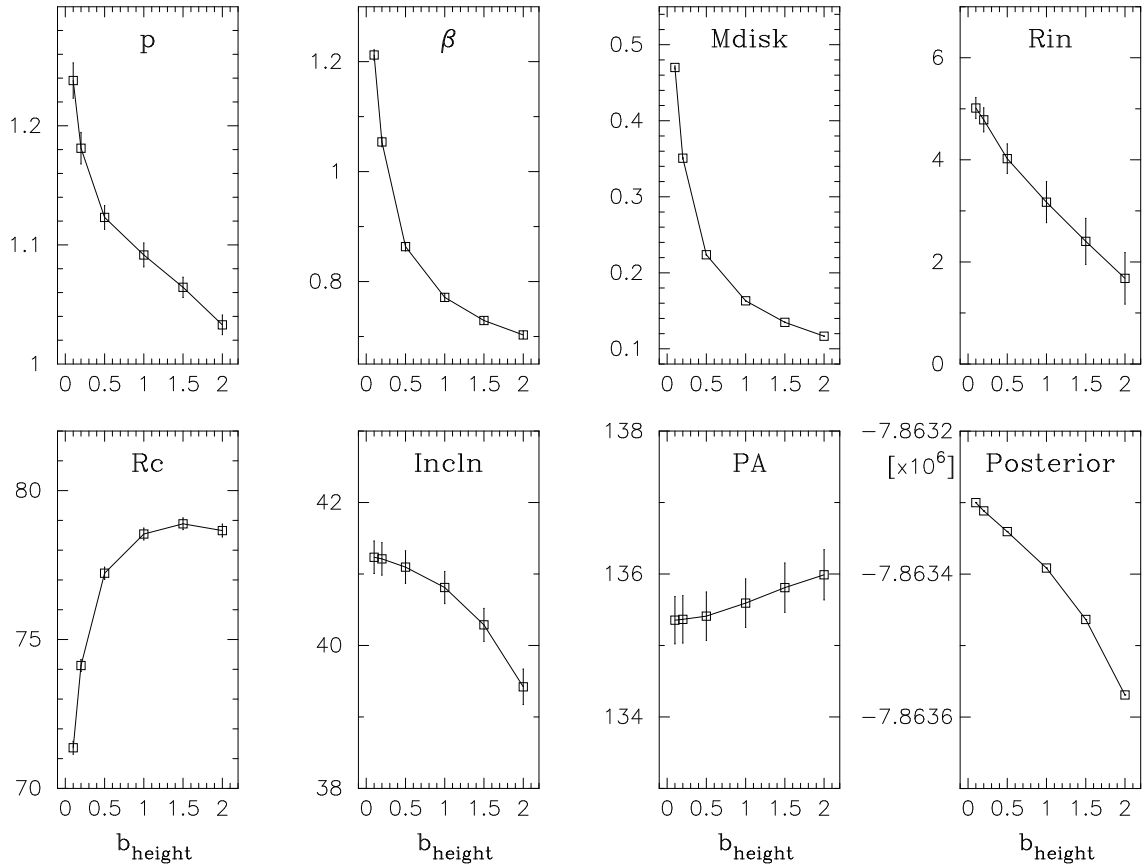}
\caption{
Free parameter variation along changing $b_{height}$.
The thinner models are preferred by our two wavelength
images, based on the posterior values.
However, thin disk models 
cannot explain the mid-infrared fluxes (Figure \ref{fig_SED}).
\label{fig_diff_b}}
\end{figure}

\begin{figure}
\includegraphics[angle=270,scale=0.6]{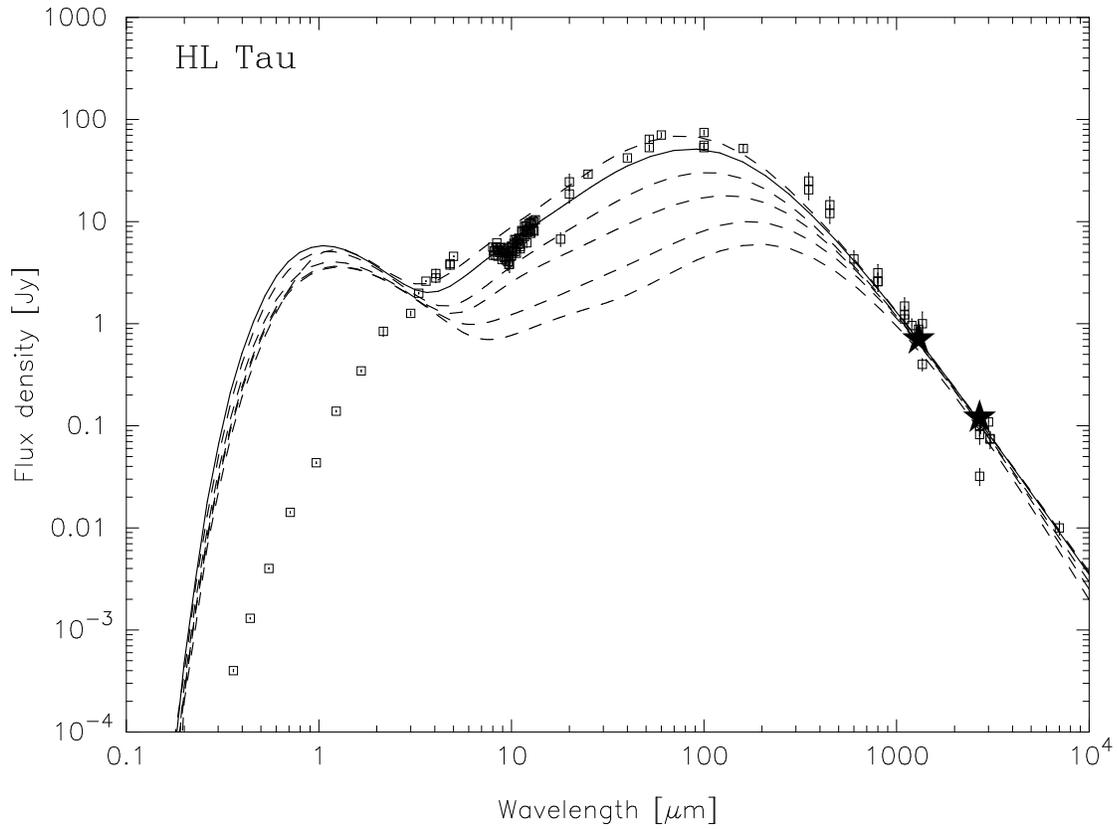}
\caption{
HL Tau SED overlaid with models of various $b_{height}$ values.
The solid line is the case of $b_{height} = 1.5$ and
the dashed lines are cases of $b_{height} = 0.1$, 0.2, 0.5, 
1.0, and 2.0 from the bottom.
The two stars indicate our data points 700 and 120 mJy 
at $\lambda=1.3$ and 2.7 mm, respectively.
\label{fig_SED}}
\end{figure}

\begin{figure}
\includegraphics[angle=270,scale=0.8]{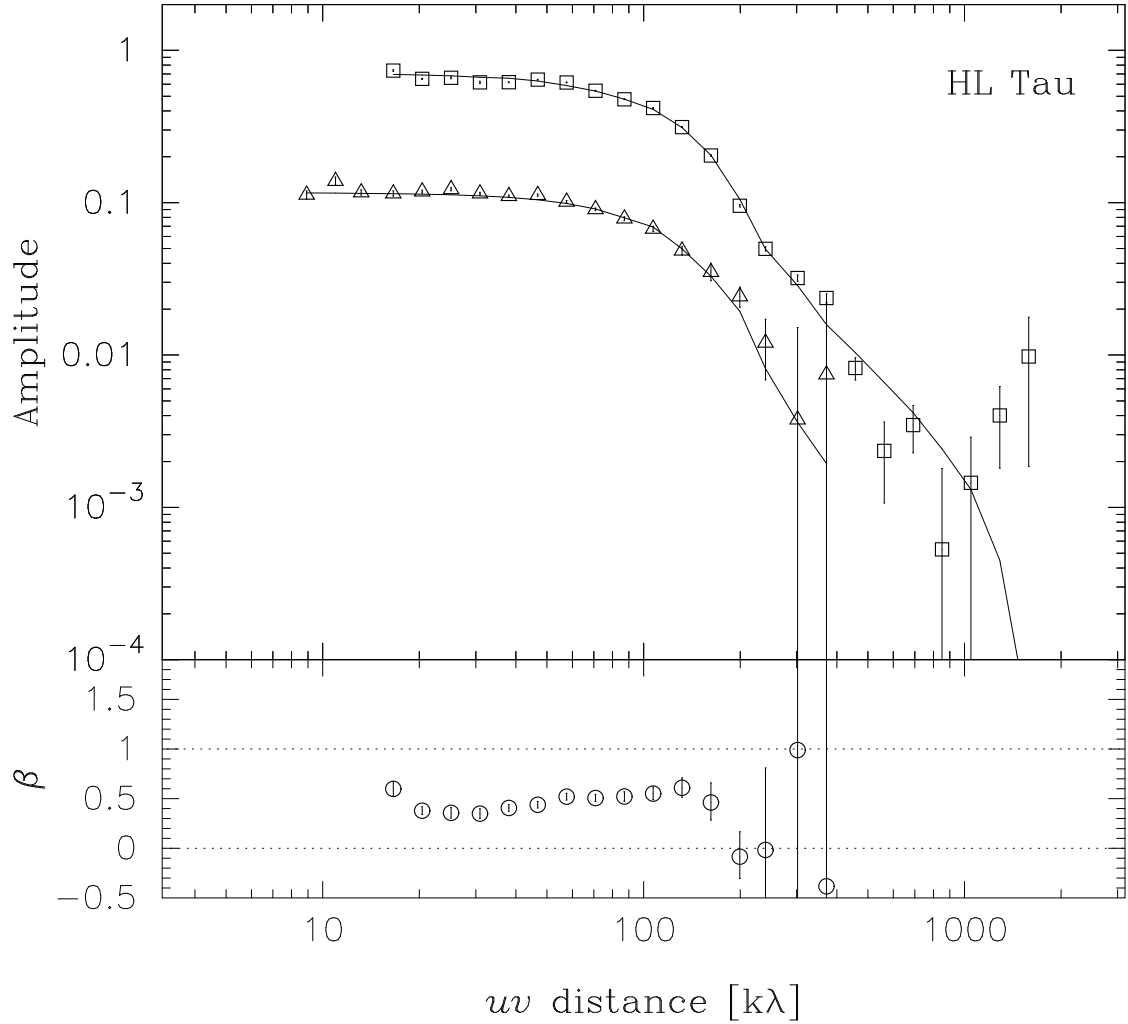}
\caption{
HL Tau visibility averaged in annuli.
The rectangular and triangle points are data at \omm\ and 2.7 mm,
respectively, and
the solid line is the model best fitting to both images and SED.
\label{fig_uvamp}}
\end{figure}

\begin{figure}
\includegraphics[angle=270,scale=0.6]{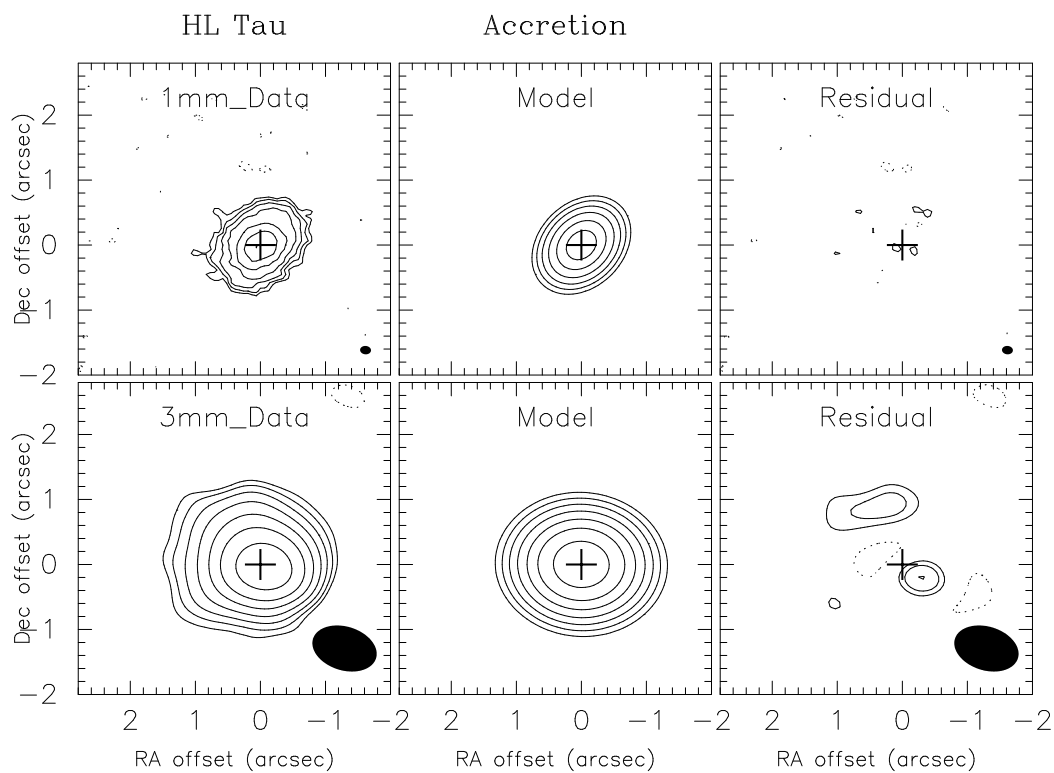}
\caption{
HL Tau continuum, model, and residual maps at \omm\ and 2.7 mm.
The \omm\ image is the same as Figure \ref{fig_hltau} and the
synthesized beam of the \tmm\ image is $0\farcs 98 \times 0\farcs
70$ ($PA = 80 \degr$).  The contour levels are 2.5, 4.0, 6.3, 10, 16, 25,
and 40 times $\sigma=\pm0.8$ and $\pm1.1$ mJy beam$^{-1}$ at \omm\ and 2.7
mm, respectively.
\label{fig_OMR}}
\end{figure}

\begin{figure}
\includegraphics[angle=270,scale=0.6]{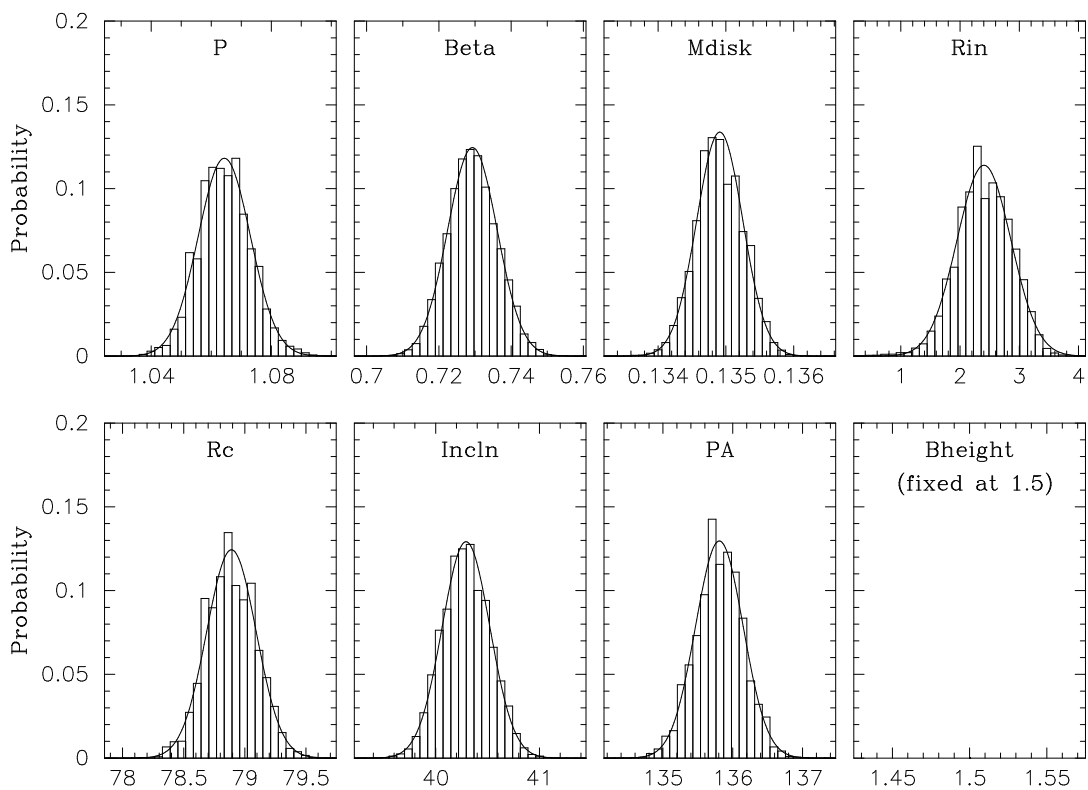}
\caption{
Posterior distributions of parameters in HL Tau.
The solid lines of the plots indicate normal distributions of the
posterior-weighted means and standard deviations of the parameter
distributions.  The posterior-weighted means and standard deviations
are listed in Table \ref{tab_parampost} with the parameter search
ranges.
\label{fig_posthl}}
\end{figure}

\begin{figure}
\includegraphics[angle=0,scale=0.8]{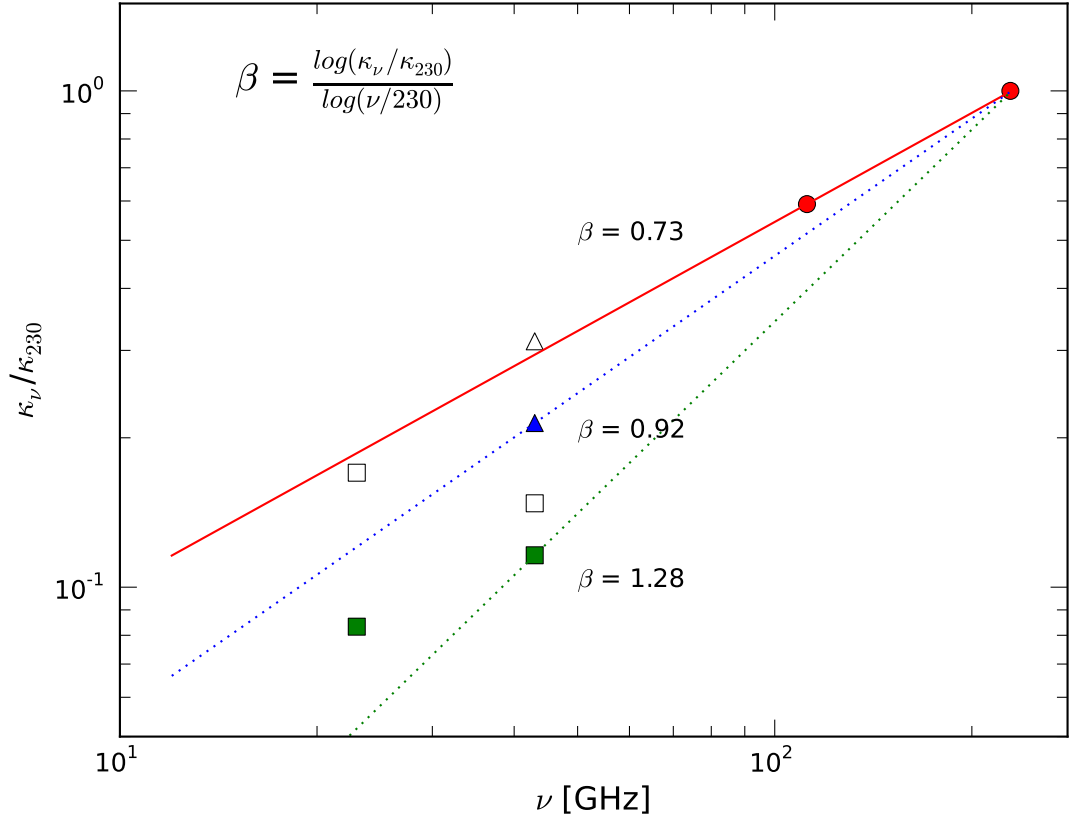}
\caption{
Mass absorption coefficients ($\kappa_\nu$) to produce the measured 
fluxes at lower frequencies with our disk model. 
The slopes present dust opacity spectral indexes ($\beta$)
determined between a frequency and $\nu = 230$ GHz.
The circles mark our results, the triangles
are of \citet{1996ApJ...470L.117W}, and the
squares are of \citet{2006A&A...446..211R}.
The open and filled symbols present the mass absorption
coefficients when
free-free emission components are included and excluded, respectively.
\citet{1996ApJ...470L.117W} estimated 30\% of their flux
$10 \pm 1.4$ mJy at $\lambda = 7$ mm for free-free emission and
\citet{2006A&A...446..211R} argued 20\% of their flux 4.92 mJy
at $\lambda = 7$ mm and 50\% of 1.63 mJy at $\lambda = 1.3$ cm
for free-free emission. 
\label{fig_kappa}}
\end{figure}

\begin{figure}
\includegraphics[angle=270,scale=0.8]{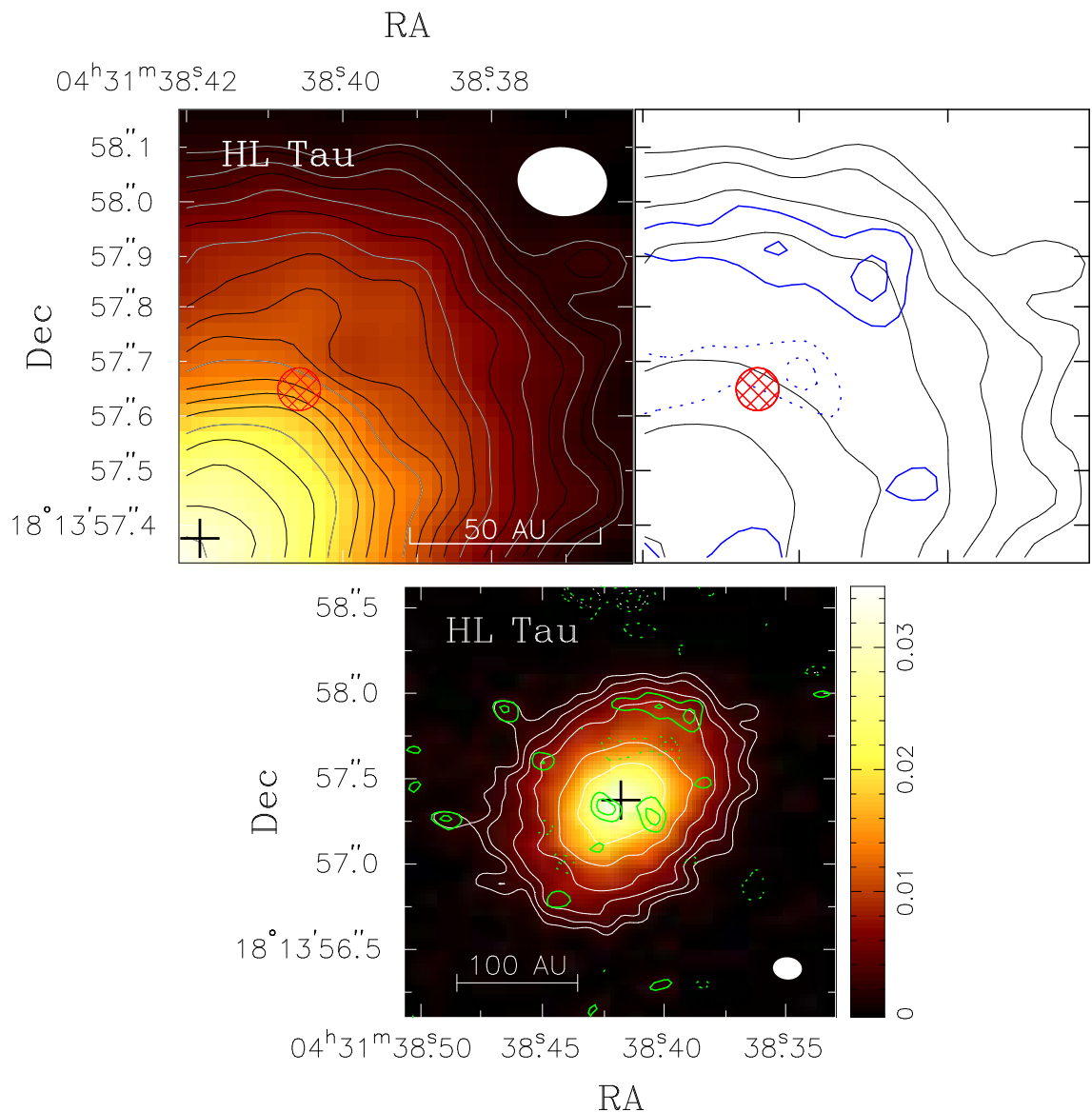}
\caption{
HL Tau in the \omm\ continuum overlaid with residual contours
in the bottom.
The disk image is the same as Figure \ref{fig_hltau} and the levels
of residual contours are 2 and 3 times $\sigma=\pm0.8$ mJy beam$^{-1}$.
The top row shows the northwest quadrant region zoomed-in.
The contour levels of the left are
2.5, 3.0, 4.0, 5.0, 6.3, 8.0, 10.0, 11.5, 13.0, 14.5, 16.0, 18.0, 20.0,
22.0, 25.0, 28.0, 32.0, 36.0, and 40.0 times $\sigma=\pm0.8$ mJy beam$^{-1}$.
The red hashed circles indicate where a protoplanet has been claimed.
As shown, no excess signal has been detected.
\label{fig_hltau_wres}}
\end{figure}

\begin{figure}
\includegraphics[angle=270,scale=0.6]{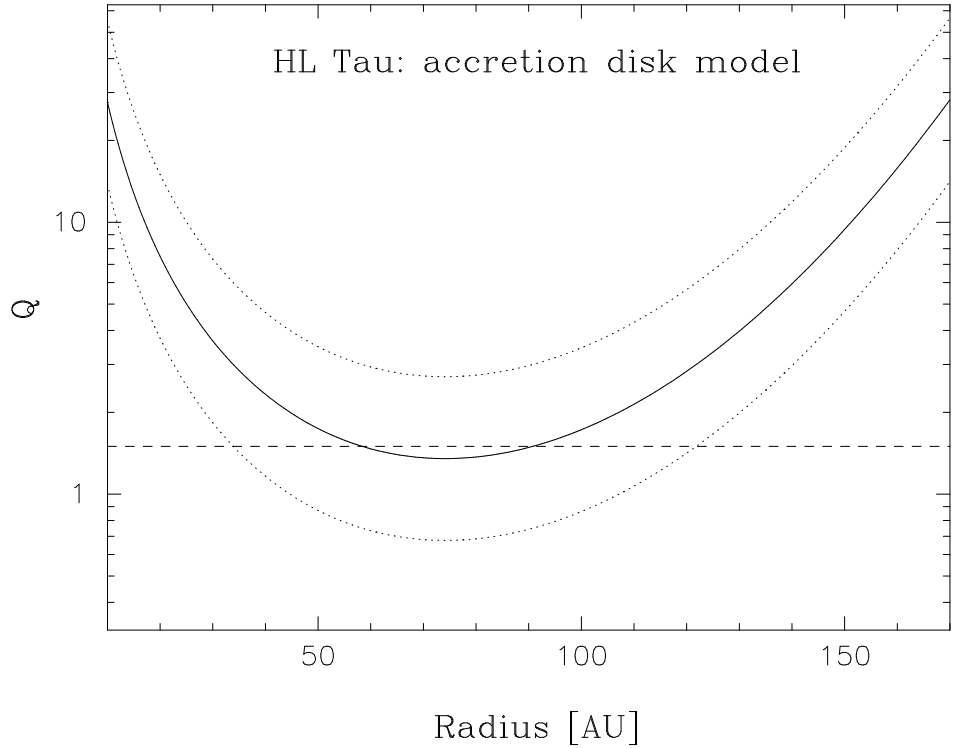}
\caption{
Toomre $Q$ parameter along radius in
HL Tau.  The dotted lines indicate the region when assuming factor
of two uncertainty in the disk mass mainly induced by $\kappa_0$
uncertainty \citep{ossenkopf1994}.  Note that the Q value is smaller than
or very close to 1.5 between 50 and 100 AU, which suggests that the
region may be gravitationally unstable.
\label{fig_qplot}}
\end{figure}

%% file: mstables.tex
\begin{table}
\caption[Disk fitting results]{Disk fitting results
of free parameters.  
\label{tab_parampost}}
\begin{tabular}{llr@{}lccc}
\hline \hline
\multicolumn{2}{c}{Parameters} & \multicolumn{2}{c}{Search ranges} &
Fitting results &  Statistical & Systematic \\
 & & & & & \multicolumn{2}{c}{Uncertainties}\\
\hline
$p$ & & 0.00&--3.35 & 1.064& 0.0085 & 0.07\tablenotemark{a} \\
$\beta$ & & -0.35&--1.7 & 0.729& 0.0069 & 0.25\tablenotemark{b} \\
$M_{disk}$ & [\msun] & 0.005&--0.2 & 0.1349& 0.00034 & 0.013\tablenotemark{c} \\
$R_{in}$ & [AU] & 0.1&--20 & 2.4& 0.45  & 2.1\tablenotemark{d}\\
$R_{c}$ & [AU] & 40&--460 & 78.9& 0.20  & 9\tablenotemark{e}\\
$\theta_i$ & [$\degr$] & 0&--83 & 40.3& 0.23  & \\
PA & [$\degr$] & 0&--180 & 135.8& 0.35  & \\
\hline
$b_{height}$ &  & 0.05&--2.0 & 1.5& \multicolumn{2}{c}{fixed} \\
\hline
$\gamma$ & & & & -0.22 & \multicolumn{2}{c}{derived from} \\
$R_t$ & [AU] & & & 40.3 & \multicolumn{2}{c}{fitted parameters} \\
\hline
\end{tabular}
\tablenotetext{a}{
Roughly estimated assuming that A-configuration data have 
a 10\% flux calibration error with respect to the other configuration 
data.  Note that relative errors in flux calibration over different
configurations can cause a gradient change along radius in brightness.
The $p$ is dependent on the assumed functional description in 
Equation (\ref{eq_rho}) and the assumed value of $q$.}
\tablenotetext{b}{Estimated from the assumed 10\% and
8\% uncertainties in the \omm\ and \tmm\ fluxes, respectively.}
\tablenotetext{c}{Estimated using the assumed flux calibration 
uncertainty in the \omm\ flux.  Note that the disk mass is sensitive 
to the temperature scale ($T_0$), the dust mass absorption coefficient, 
and the gas-to-dust ratio.  Therefore, the total uncertainty in the 
value can be up to about a factor of two.}
\tablenotetext{d}{Estimated from mapped model images with a varying 
inner edge of the disk to give a detectable emission decrease in the 
central position. }
\tablenotetext{e}{Estimated from half of the angular resolution.}
\end{table}